# $2^{nd}$ UML 2 Semantics Symposium: Formal Semantics for UML

Manfred Broy[1], Michelle L. Crane[2], Juergen Dingel[2],
Alan Hartman[3], Bernhard Rumpe[4], and Bran Selic[5]

[1] Technische Universität München, Germany
[2] Queen's University, Kingston, Ontario, Canada
[3] IBM Research, Israel
[4] Technische Universität Braunschweig, Germany
[5] IBM Rational Software, Canada
broy@in.tum.de,crane@cs.queensu.ca,dingel@cs.queensu.ca,
hartman@il.ibm.com,b.rumpe@tu-bs.de,bselic@ca.ibm.com
http://www.cs.queensu.ca/~stl/internal/uml2

**Abstract.** The purpose of this symposium, held in conjunction with MoDELS 2006, was to present the current state of research of the UML 2 Semantics Project. Equally important to receiving feedback from an audience of experts was the opportunity to invite researchers in the field to discuss their own work related to a formal semantics for the Unified Modeling Language. This symposium is a follow-on to our first workshop, held in conjunction with ECMDA 2005.

## 1 Introduction

The UML 2 Semantics Project is an international collaboration, involving both academia and industry. Participants include IBM (Canada, Germany, and Israel), Queen's University (Kingston, Ontario, Canada), the Technical University of Munich (Germany), and the Technical University of Braunschweig (Germany). The main objective of this project is to develop a mathematically formalized semantics definition for the Unified Modeling Language (UML). The Project started in January 2005 and has achieved substantial results. That said, there is much work to be done and the project will likely continue for at least one more year.

The purpose of this symposium, held in conjunction with MoDELS 2006, was to present the current state of our research to an audience of experts. Equally important to receiving feedback on our research, this symposium was an opportunity to invite researchers in the field to discuss their own work. This symposium is a follow-on to our first workshop, held in conjunction with ECMDA 2005, in Nuremberg, November 2005.

## 2   Motivation

UML has become the language of choice for modeling various aspects of software systems in academia and industry. UML is now widely adopted in academia and industry and has established itself as the dominant language for modeling software systems. UML 2 [12] is the latest major revision of UML and has been developed with the help of researchers and practitioners from numerous companies, universities, and government institutions. UML 2 addresses the shortcomings of the previous version and incorporates the advances distilled from a large body of research and practical experience. The current version of the standard specifically supports model-driven development (MDD), an approach to software development that has the proven potential to increase the productivity of industrial software development substantially. In short, MDD focuses on the construction of platform-independent, high-level models from which source code is automatically generated.

The current UML 2 specification is complex and uses a combination of semi-formal diagrams, constraints, and informal natural language text. The imprecisions and ambiguities of natural language make it difficult to detect and correct subtle errors, incompleteness, and inconsistencies. These problems in turn complicate the development of tools supporting UML. For instance, tool builders may not find the amount of detail in the standard necessary, for the implementation of a particular analysis or translation. In addition, the interoperability between UML tools is compromised, because different tools may interpret the same artifact differently, such that the combined use of these tools may not yield consistent results. The high-level goal of this project is to overcome such problems, and to improve the standard and enhance the technical viability and benefits of MDD and UML.

The proposed formalization of UML will have several benefits. First, it will allow subtle errors in the current and future versions of the standard to be detected and suggestions for improvements to be made. Second, the formalization will have the potential to be of immediate, commercial utility to the companies developing tools supporting UML and MDD. For instance, it would enable tool vendors to develop tools that offer more powerful and effective testing, analysis, and model transformation functionality and better support the exchange of modeling artifacts between different tools.

## 3   The Semantics Architecture

The focus of the Project has been driven primarily by the concepts discussed in [13], especially the *semantics architecture*. Figure 1 identifies the key semantics areas covered by the current UML 2 standard.

At the highest level of abstraction, it is possible to distinguish three distinct layers of semantics. The foundation layer is structural, reflecting the premise that there is no disembodied behaviour in UML – all behaviour emanates from the actions of structural entities. This structural layer is represented by our *System Model*, discussed in Section 4.

The next layer is behavioural and provides the foundation for the semantic description of all higher-level behavioural formalisms. This layer is called the Behavioural Base and consists of three separate sub-areas arranged into two sub-layers. The bottom sub-layer consists of the inter-object behaviour base, which deals with how structural entities communicate with each other, and the intra-object behaviour base, which the relationship between structural entities (e.g., objects) and their behaviour. The system model also formalizes these concepts. The actions sub-layer is placed over these two; it defines the semantics of individual actions and the means by which actions are composed to form more complex behavioural specifications. Actions are the fundamental units of behaviour in UML and are used to define fine-grained behaviour. As discussed in Section 5, one current document in the project is dedicated to formalizing these actions in terms of the system model.

Actions are available to any of the higher-level formalisms to be used for describing detailed behaviours. The topmost layer in the semantics hierarchy defines the semantics of the higher-level behavioural formalisms of UML: activities, state machines, and interactions. These formalisms are dependent on the semantics provided by the lower layers. Currently, research is being done on formalizing activities and interactions in terms of the system model.

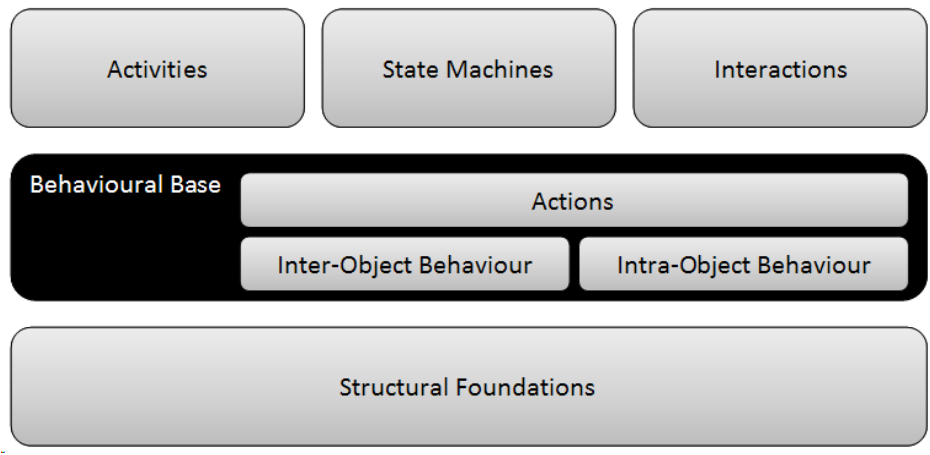

**Fig. 1.** The UML semantics layers: the Semantics Foundation consists of the bottom two layers – the Structural Foundations and the Behavioural Base [13]

## 4 System Model

The goal of the *System Model* is to provide a semantic domain into which UML specifications can be mapped [10]. In our case, the semantic domain is mathematics, specifically: numbers, sets, relations and functions. The notation is drawn from pure mathematics, as opposed to some other specialized, or invented, notation.

The system model defines a universe of interacting state machines that describe the behaviour of objects and their relationships with each other. It provides the means to define the semantics of any UML model. Intuitively, each state in the system model is composed of three parts, data store, control store and event store, and represents the states that the system being modelled moves through during its execution. Further information about the system model is detailed in the various documents listed in Section 5.

## 5 Status

Several major objectives were determined at the outset of the project:

1. To specify a definitive and complete formal semantics foundation for the UML 2 standard. At this point, approximately two-thirds of the semantics foundation has been finalized. This foundation, called the *System Model* is composed as follows:

   - Towards a System Model for UML: The Structural Data Model [3], which defines the structure part of the system model, including concepts such as class, reference, method, etc.
   - Towards a System Model for UML: The Control and Scheduling [2], which defines the control part of the system model, including concepts such as stack, frame, thread, message, etc.
   - Towards a System Model for UML: The State Transition System, which defines the dynamic behaviour of the system model.

   These three documents introduce a system model as the basis for a semantic model for UML 2. The system model forms the core and foundation of the UML semantics definition. Building upon this system model are several other documents:

   - Class Diagrams: Abstract Syntax and Mapping to System Model [5], which expresses a subset of UML class diagrams in terms of a tuple notation and then maps this structure to the system model.
   - Activity Diagrams: Abstract Syntax and Mapping to System Model [7], which expresses a subset of UML activity diagrams in terms of a tuple notation and then maps this structure to the system model.
   - Mapping Actions to the System Model [4], which examines several of the UML "primitive" actions, such as *CreateObjectAction*, *CallOperationAction*, etc. The behaviour of these actions is expressed in terms of changes to the system model.
   - Mapping Activities to the System Model [6], which examines the fundamental nature of activities, e.g., tokens, flow, how activities can be composed, etc. The result of activity execution is expressed in terms of the system model.

   At this point in time, the documents listed above are available in unpublished format only. The most current version of each document may be found online [1].
2. To identify potential consistency flaws in the UML 2 standard and propose adequate corrections. Several subtle inconsistencies and flaws in the standard have been found over the past 18 months - these have been forwarded to the appropriate authors, who have raised issues when appropriate.
3. To identify analysis techniques that can be used to formally determine the correctness of UML 2 models. These techniques would enable tool vendors to develop tools that offer more powerful and effective testing, analysis, and model transformation functionality and better support the exchange of modeling artifacts between different tools. To date, that has been little progress on this objective, although it remains a high priority for future work.
4. To provide a strong foundation for the definition of a UML virtual machine that is capable of executing UML 2 models. Progress on this objective is being made on two fronts:
   - Dr. Alan Hartman's group at IBM Haifa, Israel has created a generic model execution engine [11] on top of which a UML simulator for activity diagrams and state machines has been implemented. The simulator allows modellers to step through their models in an interactive fashion and thus gain a better understanding of their behaviour.
   - Simultaneously, research is carried out to use the system model as the basis of an execution and analysis engine. The goal of this work is to refine the system model and to pave the way towards a more powerful analysis platform based directly on our formal semantics of UML.
   
   Cross-pollination between these two initiatives is expected to benefit both.

In addition to these primary objectives, research has been conducted on these related topics: clarification of complicated or new aspects of UML, e.g., associations [8], package merge [15,14], and generic model management [9].

## 6  Future Work

The original project mandate was for two years. We have made significant progress in that period of time. Although there is much more research to be done, we are anticipating the continuation of this project for at least one more year. Regardless, the majority of the system model is nearly complete and can be used in future research. More specifically, work on mapping the actions and activities to the system model will be continued.

**Acknowledgments.** This research is supported by Communications and Information Technology Ontario, the IBM Centers for Advanced Studies, the Technische Universität München, and IBM Germany.